\documentclass[fleqn,usenatbib,useAMS]{mnras}
\usepackage{amssymb,amsmath}
\usepackage{graphicx}
\usepackage{xcolor,natbib}
\usepackage{multirow}
\usepackage[T1]{fontenc}
\usepackage{ae,aecompl}
\usepackage{newtxtext, newtxmath}
\usepackage{float}
\usepackage{subfigure}
\usepackage{longtable}
\usepackage{enumitem}
\usepackage{longtable,listings}
\usepackage[flushleft]{threeparttable}
\usepackage{parcolumns}
\def\obj{SDSS J2219-0938}

\title[Dual core system]{Evidence to disfavour dual core system leading to double-peaked narrow emission lines}
\author[Zhang \& Zheng]{XueGuang Zhang$^{1}$
\thanks{Contact e-mail: \href{mailto:xgzhang@gxu.edu.cn}{xgzhang@gxu.edu.cn}}, Qi Zheng$^{2}$\\
$^{1}$Guangxi Key Laboratory for Relativistic Astrophysics, School of Physical Science and Technology, 
	Guangxi University, Nanning, 530004, P. R. China\\
$^{2}$School of Physics and Technology, Nanjing Normal University, No. 1, Wenyuan Road, Nanjing, 
	230046, P. R. China}

\begin{document}
\label{firstpage}
\pagerange{\pageref{firstpage}--\pageref{lastpage}}

\maketitle

\begin{abstract} %%%about 190 words
	In this manuscript, an interesting method is proposed to test dual core system for double-peaked narrow 
emission lines, through precious dual core system with double-peaked narrow Balmer lines in one system in 
main galaxy but with single-peaked narrow Balmer lines in the other system in companion galaxy. Under a dual 
core system, considering narrow Balmer (H$\alpha$ and H$\beta$) emissions ($f_{e,~\alpha}$ and $f_{e,~\beta}$) 
from companion galaxy but covered by SDSS fiber for the main galaxy and narrow Balmer emissions 
($f_{c,~\alpha}$ and $f_{c,~\beta}$) from the companion galaxy covered by SDSS fiber for the companion galaxy, 
the same flux ratios $f_{e,~\alpha}/f_{c,~\alpha}=f_{e,~\beta}/f_{c,~\beta}$ can be expected, due to 
totally similar physical conditions of each narrow Balmer emission region. Then, the precious dual core system 
in SDSS J2219-0938 is discussed. After subtracting pPXF code determined stellar lights, double-peaked narrow 
Balmer emission lines are confirmed in the main galaxy with confidence level higher than $5\sigma$, but 
single-peaked narrow Balmer emission lines in the companion galaxy. Through measured fluxes of emission 
components, $f_{e,~\alpha}/f_{c,~\alpha}$ is around 0.82, different from $f_{e,~\beta}/f_{c,~\beta}\sim0.52$, 
to disfavour a dual core system for the double-peaked narrow Balmer emission lines in SDSS J2219-0938.
\end{abstract}


\section{Introduction}

%%first
	Dual core systems on scale of dozens to thousands parsecs to supermassive binary black hole (BBH) systems 
on scale of sub-parsecs in galaxies are commonly expected products by merging of galaxies, essential process of 
galaxy formation and evolution \citep{bb80, kw93, sr98, md06, mk10, rs17, bh19, mj21, mj22, yp22}. And there are 
many techniques proposed to detect dual core systems and BBH systems. Applications of double-peaked features of 
broad and/or narrow emission lines can be found in \citet{kz08, bl09, sl10, pl12, cs13, le16, dv19}. Applications 
of spatially resolved image properties of central regions have been reported in \citet{km03, rt09, pv10, ne17, kw20}. 
Applications of different line profiles of broad Balmer emission lines can be found in \citet{zh21d}. Applications 
of long-standing Optical Quasi-Periodic Oscillations have been reported in \citet{gd15, kp19, ss20, lw21, zh22}.

\begin{figure*}
\centering\includegraphics[width = 18cm,height=7.75cm]{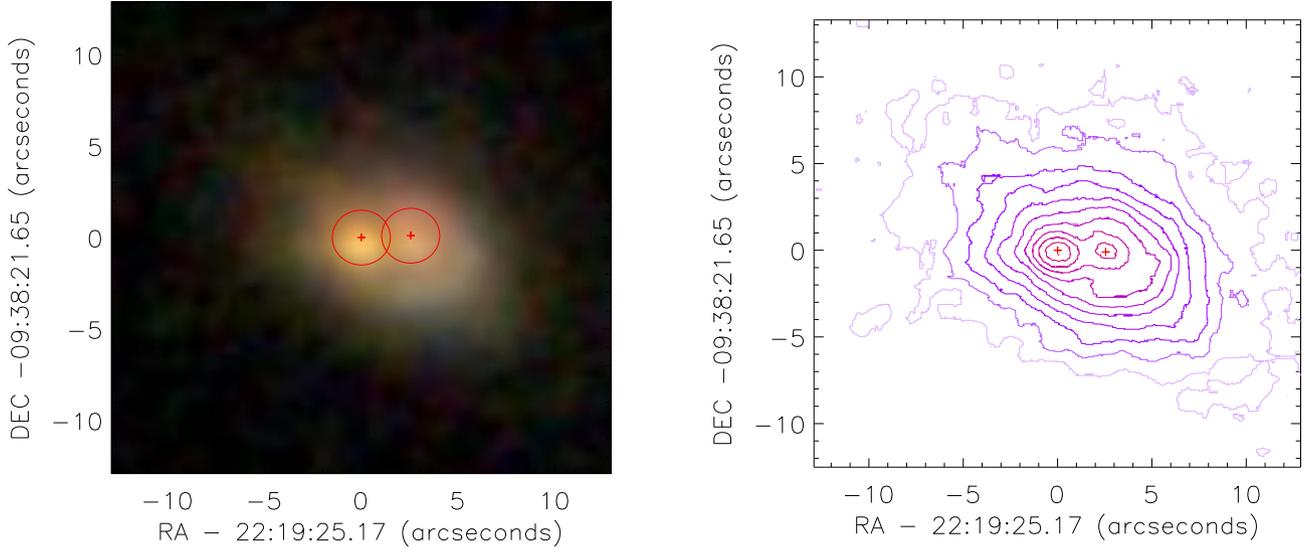}
\caption{Left panel shows the 25.8\arcsec$\times$25.8\arcsec~ colorful photometric image of \obj, right panel 
shows corresponding brightness contour of the photometric image. In left panel, the two red circles with radii 
1.5\arcsec~ represent covering regions of the SDSS fibers, the two red pluses represents the pointing positions 
(central positions of the two nuclei) of the SDSS fibers. In right panel, the two red pluses mark the central 
positions of the two nuclei.}
\label{img}
\end{figure*}

	Among the proposed techniques to detect dual core systems, application of double-peaked narrow emission 
lines is a very interesting topic. \citet{zw04} have firstly reported a dual core system in SDSS J1048+0055 
through double-peaked narrow emission lines combining with radio properties. And then, \citet{gn07} have reported 
a dual core system in EGSD2 J1420+5259 through double-peaked [O~{\sc iii}] emission features combining with 
multi-band wavelength properties. \citet{xk09} have reported a dual core system in SDSS J1316+1753 through all 
narrow emission lines having double-peaked features. \citet{lg10} have reported dual core systems in four objects 
through double-peaked [O~{\sc iii}] emission features combining properties of deep near-infrared images. 
\citet{mm11} have reported a dual core system in SDSS J0952+2552 through its double-peaked [O~{\sc iii}] lines 
combining with properties of resolved near-infrared images. \citet{fz11} have reported a dual core system in 
SDSS J1502+1115 through double-peaked [O~{\sc iii}] emission features and resolved radio images. \citet{bs12} 
have shown a dual core system in CXOXB J1426+3533 through double-peaked narrow emission line features combining 
with properties of near-infrared adaptive optics imaging. \citet{bs13} have shown that the dual core systems are 
favoured to the double-peaked high-ionization narrow emission lines through a sample of 131 quasars with $0.8<z<1.6$. 
\citet{wc14} have reported a dual core system through double-peaked narrow emission lines combining with Hubble 
Space Telescope imaging. \citet{sb21} have reported a favoured dual core system in SDSS J1431+4358 with 
double-peaked narrow emission lines. Besides the discussed individual objects, there are large samples of objects 
with double-peaked narrow emission lines reported in \citet{ss10, gh12, wl19}, etc..

	However, there are some further reports to disfavour the double-peaked narrow emission lines as efficient 
signs of dual core systems. \citet{ls10} have shown double-peaked features due to narrow-line region kinematics 
or dual core systems. \citet{rs10} have shown double-peaked narrow emission lines due to radio-jet driven outflows. 
\citet{fm11} have shown scenarios involving a single AGN leading to the same double-peaked narrow emission lines. 
\citet{sl11} have shown kinematics scenario with a single AGN can be commonly applied for the majority of 
double-peaked [O~{\sc iii}] lines in Type-2 AGN. \citet{fy12} have discussed that only probably 1\% dual AGN can 
lead to double-peaked narrow emission lines. \citet{zh15} have reported non-kinematic model for double-peaked 
narrow H$\alpha$. \citet{mm15} have shown that only one dual core system is detected among 12 candidates with 
double-peaked narrow emission lines, followed by \citet{mc15, nc16}. \citet{zh16} have shown that dual core systems 
are not statistically preferred to double-peaked narrow emission lines, through virial BH mass comparisons of 
broad line AGN with and without double-peaked narrow emission lines. \citet{ll18} have shown that radio-loud 
double-peaked narrow emission line AGN should be related to jets.

	In the manuscript, an interesting method is proposed to test the dual core system for the double-peaked narrow 
emission lines. Section 2 presents our main hypotheses. Section 3 presents main results and necessary discussions 
in the precious dual core system in SDSS J221924.98-093821.6 (=SDSS J2219-0938). Section 4 gives final conclusions. 
And, the cosmological parameters have been adopted as $H_{0}=70{\rm km\cdot s}^{-1}{\rm Mpc}^{-1}$, 
$\Omega_{\Lambda}=0.7$ and $\Omega_{\rm m}=0.3$.

\section{Main Hypotheses}

	The main considering point is that narrow Balmer emission line regions covered by SDSS fiber have totally 
the same physical conditions. Then, a kind of precious dual core system, one core with double-peaked narrow Balmer 
emission lines in the main galaxy but the other core with single-peaked narrow Balmer emission lines in the 
companion galaxy, can be well discussed.

	For the double-peaked narrow Balmer emission lines (H$\alpha$ and H$\beta$) in the main galaxy, the 
blue-shifted (or red-shifted) components of the double-peaked narrow Balmer lines definitely include contributions 
of the narrow Balmer emissions from the companion galaxy but covered by the fiber for the main galaxy, under the 
framework of a dual core system. In other words, from the intrinsic narrow Balmer emission line regions of the 
companion galaxy, one region (ext-emission region) is covered by the SDSS fiber for the main galaxy, the other one 
region (comp-emission region) is covered by the SDSS fiber for the companion galaxy. Considering $p1$ and $p2$ 
as the ratio of narrow Balmer emissions in ext-emission region and in comp-emission region to the intrinsic total 
narrow Balmer emissions in the companion galaxy, we will have 
\begin{equation}
%\begin{split}
\frac{f_{e,~\alpha}}{f_{T,~\alpha}}~=~p1,~~\frac{f_{e,~\beta}}{f_{T,~\beta}} = p1,~~
	\frac{f_{c,~\alpha}}{f_{T,~\alpha}}~=~p2,~~\frac{f_{c,~\beta}}{f_{T,~\beta}} = p2
%\end{split}
\end{equation}
with $f_{e,~\alpha}$ and $f_{e,~\beta}$ ($f_{c,~\alpha}$ and $f_{c,~\beta}$) as line fluxes of narrow H$\alpha$ 
and narrow H$\beta$ from the ext-emission region (the comp-emission region), and $f_{T,~\alpha}$ and $f_{T,~\beta}$ 
as intrinsic total line fluxes of narrow H$\alpha$ and narrow H$\beta$ of the companion galaxy. Although $p1$ and 
$p2$ are unknown parameters, the equations above can well lead to 
\begin{equation}
	\frac{p1}{p2}=\frac{f_{e,~\alpha}}{f_{c,~\alpha}}=\frac{f_{e,~\beta}}{f_{c,~\beta}}=R_{ec}
\end{equation}
Therefore, properties of $R_{ec}$ can be well applied to test the assumed dual core system for double-peaked 
narrow emission lines.

	Certainly, due to different locations of ext-emission region and comp-emission region, different dust 
obscurations should have effects on the flux ratio $R_{ec}$. However, considering intrinsic narrow Balmer 
decrement (flux ratio of narrow Balmer emission lines) can be totally applied to correct the effects of dust 
obscurations. Then, in the manuscript, an interesting target \obj~ is collected and discussed, due to its 
double-peaked narrow Balmer emission lines in the main galaxy but single-peaked narrow Balmer emission lines 
in the companion galaxy.

\section{Photometric and spectroscopic results in \obj}

\begin{figure}
\centering\includegraphics[width = 8cm,height=8cm]{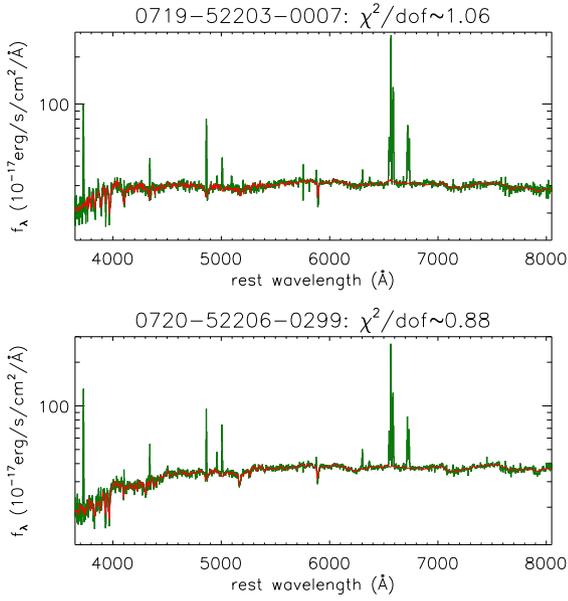}
\caption{SDSS spectra (in dark green) of the two nuclei and the pPXF code determined host galaxy contributions 
(in red). The determined $\chi^2/dof$ related to the best descriptions is marked in title of each panel.
}
\label{spec}
\end{figure}

\begin{figure}
\centering\includegraphics[width = 7cm,height=14.25cm]{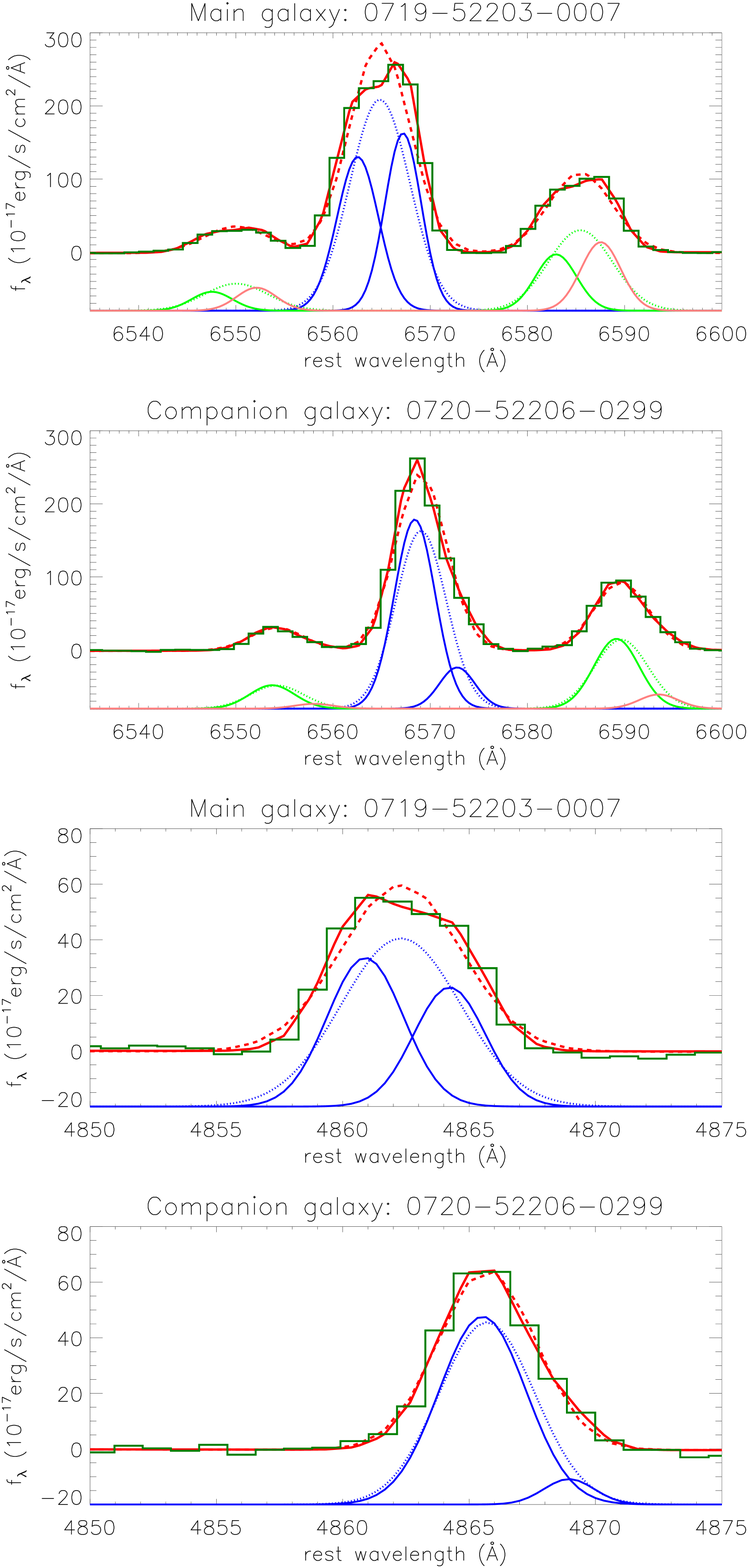}
\caption{Top two panels show the best-fitting results (solid red line for model B, dashed red line for model A) 
to the emission lines around H$\alpha$ (in dark green) in the main galaxy and in the companion galaxy. Bottom 
two panels show corresponding results to the emission lines around H$\beta$. In top two panels, dotted blue 
line and dotted green lines show the model A determined narrow H$\alpha$ and [N~{\sc ii}] doublet, solid blue 
lines, solid green lines and solid pink lines show the model B determined Gaussian components in narrow H$\alpha$ 
and [N~{\sc ii}] doublet. In bottom two panels, dotted blue line shows the model A determined narrow H$\beta$, 
solid blue lines show the model B determined Gaussian components in narrow H$\beta$.}
\label{line}
\end{figure}

\begin{table*}
\caption{Line parameters}
\begin{tabular}{lllllll|llllll}
\hline\hline
line & $\lambda_0$ & $\sigma$ & flux & $\lambda_0$ & $\sigma$ & flux & $\lambda_0$ & $\sigma$ & flux & 
	$\lambda_0$ & $\sigma$ & flux\\
	& \multicolumn{3}{c}{model A in main galaxy} &  \multicolumn{3}{c}{model B in main galaxy} &
	\multicolumn{3}{c}{model A in companion galaxy} &  \multicolumn{3}{c}{model B in companion galaxy} \\	
\hline
\multirow{2}{*}{H$\alpha$}  &   \multirow{2}{*}{6564.8$\pm$0.1}  & \multirow{2}{*}{3.1$\pm$0.1}  & \multirow{2}{*}{223$\pm$2} &
	6562.5$\pm$0.1 & 2.1$\pm$0.1 & 112$\pm$5 &
	\multirow{2}{*}{6567.8$\pm$0.1}  & \multirow{2}{*}{2.6$\pm$0.1}  & \multirow{2}{*}{159$\pm$15} &
	6567.1$\pm$0.1 & 2.1$\pm$0.1 & 136$\pm$5  \\
	&   &   &   &
	6567.2$\pm$0.1 & 1.9$\pm$0.1 & 113$\pm$5
	&   &   &  & 6571.5$\pm$0.3 & 1.8$\pm$0.2 & 26$\pm$5 \\
\hline
\multirow{2}{*}{H$\beta$}  &  \multirow{2}{*}{4862.3$\pm$0.1}  & \multirow{2}{*}{2.4$\pm$0.1}  & \multirow{2}{*}{36$\pm$7} &
	4860.9$\pm$0.2 & 1.5$\pm$0.1 & 20$\pm$3 &
	\multirow{2}{*}{4864.7$\pm$0.1}  & \multirow{2}{*}{1.9$\pm$0.1}  & \multirow{2}{*}{31$\pm$7} &
	4864.5$\pm$0.1 & 1.7$\pm$0.1 & 29$\pm$2  \\
	&   &   &   &
	4864.2$\pm$0.2 & 1.4$\pm$0.1 & 15$\pm$3
	&   &   &  & 4867.9$\pm$0.4 & 1.0$\pm$0.4 & 2$\pm$1\\
\hline
\multirow{2}{*}{[N~{\sc ii}]}  &  \multirow{2}{*}{6585.4$\pm$0.1}  & \multirow{2}{*}{3.3$\pm$0.1}  & \multirow{2}{*}{91$\pm$10} &
	6582.9$\pm$0.2 & 2.1$\pm$0.1 & 41$\pm$4 &
	\multirow{2}{*}{6588.3$\pm$0.2}  & \multirow{2}{*}{2.8$\pm$0.1}  & \multirow{2}{*}{65$\pm$9} &
	6587.8$\pm$0.2 & 2.3$\pm$0.1 & 56$\pm$6 \\
	&   &   &   &
	6587.5$\pm$0.2 & 2.1$\pm$0.1 & 48$\pm$4
	&   &   &  & 6592.2$\pm$0.9 & 2.0$\pm$0.4 & 9$\pm$5 \\
\hline\hline
\end{tabular}\\
Notice: the center wavelength $\lambda_0$ in unit of \AA, the line width (second moment) $\sigma$ in unit of \AA~ 
and the line flux in unit of ${\rm 10^{-16}~erg/s/cm^2}$. \\
\end{table*}

%1
	\obj~ at redshift 0.0948 is firstly reported with double-peaked narrow emission lines in \citet{gh12}. 
Moreover, there are not only double-peaked narrow emission lines and apparent dual core in photometric image, 
but also there are SDSS providing high quality spectroscopic results of the dual core, leading to the best 
chance to test dual core system applied to explain the double-peaked narrow emission lines in \obj, through 
properties of $R_{ec}$ discussed above.

	Left panel of Fig.~\ref{img} shows the 25.8\arcsec$\times$25.8\arcsec~ colorful photometric image with 
central position of RA=22:19:24.98 and DEC=-09:38:21.6. Right panel of Fig.~\ref{img} shows the image in contour, 
leading to two apparent cores with central positions (brightness peak positions) of (RA=334.8548,~DEC=-9.63938) 
and (RA=334.8541,~DEC=-9.63932) marked as red pluses, leading to the projected space distance about 2.6\arcsec~ 
(5480pc at redshift 0.0948) between the two nuclei. Meanwhile, the two nuclei have SDSS spectra with plate-mjd-fiberid 
to be 0720-52206-0299 (covered by the fiber for the companion galaxy) and 0719-52203-0007 (for the main galaxy). 
The SDSS fiber covered areas are marked as red circles in left panel of Fig.~\ref{img}, with the central positions 
of the two circles as the pointing positions (totally similar as the central positions of the two nuclei) of SDSS 
fibers.

	Fig.~\ref{spec} shows SDSS spectra of the two nuclei, with signal-to-noise about 22-25. In order to 
show clear narrow emission line features, the more recent pPXF (penalized pixel-fitting) code \citep{cm17}, 
one commonly applied SSP (Simple Stellar Population) method \citep{bc03, ka03}, is accepted to determine stellar 
contributions. The pPXF code is applied with 224 SSP templates from the MILES stellar library \citet{fs11} with 
32 stellar ages from 0.06Gyrs to 17.78Gyrs and with 7 metallicities from -2.32 to 0.22. After considering the 
popular regularization method, the pPXF code can give reliable and smoother star-formation histories. 
Fig.~\ref{spec} shows the determine stellar lights in SDSS spectra of the two nuclei. And the pPXF code 
determined shifted velocities for the stellar templates are about $-17\pm11{\rm km/s}$ and $-230\pm14{\rm km/s}$ 
to describe the stellar features in the main galaxy and in the companion galaxy, respectively. The absorption 
features in the main galaxy are accepted to determine shifted velocities of the emission lines in the main galaxy 
and in the companion galaxy.

%%%2
	After subtracting host galaxy contributions and corrected pPXF code determined shifted properties, narrow 
Balmer emission lines can be well measured by multiple Gaussian functions, similar as what we have recently done 
in \citet{zh21a, zh21b, zh22a, zh22b, zh22c}. Emission lines around H$\alpha$ with rest wavelength from 6520 to 
6620\AA~ are firstly discussed, including narrow H$\alpha$ and [N~{\sc ii}] doublet. In order to confirm the 
double-peaked narrow line emission features, two different model functions are applied. In model A, each narrow 
emission line is described by one Gaussian function. But in model B, each narrow emission line is described by 
two Gaussian functions. When the model functions in both model A and model B are applied, the following two criteria 
are accepted. First, each Gaussian component has emission flux not smaller than zero. Second, components in 
[N~{\sc ii}] doublet have the same redshift, the same line width in velocity space and have the flux ratio to 
be fixed to the theoretical value of 3. Then, through the Levenberg-Marquardt least-squares minimization technique 
(MPFIT package), narrow H$\alpha$ and [N~{\sc ii}] doublet can be well measured. The best fitting results are shown 
in top two panels of Fig.~\ref{line} with $\chi^2_A/dof_A=1026.4/58\sim17.7$, $\chi^2_B/dof_B=84.4/52\sim1.6$ and 
$\chi^2_A/dof_A= 380.8/59\sim6.5$, $\chi^2_B/dof_B=87.4/53\sim1.6$, through applications of model A and model B 
to the lines in the spectra of the main galaxy and of the companion galaxy, respectively. Parameters of the emission 
components are listed in Table~1. In order to show clearer results in Fig.~\ref{line}, the plots are limited with 
wavelength from 6530 to 6600\AA.

	Similar as what we have recently done in \citet{zh22c}, the F-test statistical technique is applied to 
determine whether the functions (double-peaked features) in model B are preferred in the main galaxy. Based on the 
different $\chi^2/dof$ values for model A and Model B, the calculated $F_p$ value is about
\begin{equation}
F_p=\frac{\frac{\chi^2_A-\chi^2_B}{dof_A-dof_B}}{\chi^2_B/dof_B}\sim97
\end{equation}
Based on $dof_A-dof_B$ and $dof_A$ as number of dofs of the F-distribution numerator and denominator, the expected 
value from the statistical F-test is only about 10 (very smaller than 97) with $5\sigma$ confidence level. Therefore, 
the confidence level is higher than $5\sigma$ to support the double-peaked narrow H$\alpha$ and [N~{\sc ii}] doublet. 
Similar procedure is applied to the model determined results in the companion galaxy, leading to higher than $5\sigma$ 
confidence level to support the model B determined results. However, as listed parameters in Table~1 for emission 
lines in the companion galaxy, there are no reliable measurements of double-peaked features in [N~{\sc ii}] doublet, 
due to measured emission intensity smaller than corresponding uncertainty. Therefore, the applied model B in the 
companion galaxy can lead to an extended component in narrow H$\alpha$, but no double-peaked features.

	Similar procedures can be applied to measure the narrow H$\beta$ within rest wavelength range from 4830 to 
4900\AA. Only one Gaussian component is applied in model A, and two Gaussian functions are applied in model B. Then, 
through the Levenberg-Marquardt least-squares minimization technique, narrow H$\beta$ can be well measured. The best 
fitting results are shown in Fig.~\ref{line} with $\chi^2_A/dof_A=124.3/57\sim2.2$,~$\chi^2_B/dof_B=50.6/54\sim0.94$ 
and $\chi^2_A/dof_A=60.7/58\sim1.0$, $\chi^2_B/dof_B=47.3/55\sim0.86$,through applications of model A and model B to 
the lines in the spectra of the main galaxy and the companion galaxy, respectively. In order to show clearer results 
in Fig.~\ref{line}, the plots are limited with wavelength from 4850 to 4875\AA. And the parameters of the emission 
components are also listed in Table~1. Furthermore, through the F-test statistical technique and the determined 
different $\chi^2/Dof$ values for the model A and model B, higher than $5\sigma$ confidence level can be determined 
to support the model B determined results in the main galaxy indicating reliable double-peaked narrow H$\beta$. 
Meanwhile, smaller than $3\sigma$ confidence level can be determined to support the model B determined results in 
the companion galaxy, moreover considering measured emission intensity smaller than 2times of corresponding uncertainty, 
model A determined results are preferred in the companion galaxy, indicating single-peaked narrow H$\beta$ similar 
as the single-peaked narrow H$\alpha$ in the companion galaxy.

	Based on the well measured double-peaked narrow Balmer emission lines in the main galaxy but single-peaked 
narrow Balmer emission lines in the companion galaxy, after considering the pPXF code determined intrinsic shifted 
velocities in stellar features, properties of $R_{ec}$ can be well checked. The $R_{ec}$ in narrow H$\alpha$ is 
about $(113\pm5)/(136\pm5)\sim0.82\pm0.07$, however the $R_{ec}$ in narrow H$\beta$ is about 
$(15\pm3)/(29\pm2)\sim0.52\pm0.14$. Quite different ratios of $R_{ec}$ in narrow H$\alpha$ and in narrow H$\beta$ 
strongly indicate interesting clues not to support the assumption that the double-peaked narrow Balmer emission 
lines in the main galaxy are tightly related to emission regions belong to the central dual core system.

	Furthermore, effects of dust obscurations can be considered as follows. The narrow Balmer emission lines 
have flux ratio of $5.1_{-1.3}^{+2.1}$ ($(159\pm15)/(31\pm7)$) in the line spectrum of the companion galaxy. The 
narrow Balmer emission lines in the red-shifted components of double-peaked narrow Balmer lines in the main 
galaxy have flux ratio of $7.5_{-1.5}^{+2.3}$ ($(113\pm5)/(15\pm3)$). The similar narrow Balmer decrements in the 
comp-emissions and in the ext-emissions strongly indicate similar effects of obscurations on properties of $R_{ec}$. 
Therefore, considering dust obscurations can lead to re-confirmed different $R_{ec}$, providing clues to disfavour 
the dual core system related two emission regions leading to the double-peaked narrow Balmer emission lines in 
the main galaxy.

\section{Conclusions}

	Rotating dual narrow emission line regions in a dual core system can be applied to explain double-peaked 
narrow emission lines. However, there are more and more reports to support that double-peaked narrow emission 
lines are not efficient indicators for dual core systems. Therefore, in this manuscript, an interesting and independent 
method is proposed to test whether the dual core system can be applied to explain double-peaked narrow emission lines 
in precious dual core systems with one system having double-peaked narrow emission lines but the other system having 
single-peaked narrow emission lines. Accepted dual core system leading to double-peaked narrow emission lines, 
based on measured narrow Balmer emissions ($f_{e,~\alpha}$, $f_{e,~\beta}$) from the emission regions of the 
companion galaxy but covered by the SDSS fiber for the main galaxy and measured narrow Balmer emissions 
($f_{c,~\alpha}$, $f_{c,~\beta}$) from the emission regions of the companion galaxy covered by the SDSS fiber for 
the companion galaxy, it will be expected that $f_{e,~\alpha}/f_{c,~\alpha}=f_{e,~\beta}/f_{c,~\beta}$. Then, the 
\obj~ (the main galaxy) is collected, due to its double-peaked narrow Balmer emission lines and single-peaked narrow 
Balmer emission lines in its companion galaxy. After well measured narrow emission lines, the double-peaked narrow 
Balmer emission lines can be confirmed in the main galaxy with confidence level higher than $5\sigma$. Moreover, 
through the measured emission components in narrow Balmer emission lines in the main galaxy and in the companion 
galaxy, the flux ratio $f_{e,~\alpha}/f_{c,~\alpha}$ is about 0.82, while the flux ratio $f_{e,~\beta}/f_{c,~\beta}$ 
is about 0.52. The results indicate that the double-peaked narrow Balmer emission lines in \obj~ are not mainly 
caused by the narrow Balmer emission line regions related to the observational dual core system. A sample of such 
precious dual core systems will be discussed in near future to provide further insights on dual core systems applied 
to explain double-peaked narrow emission lines.

\section*{Acknowledgements}
%Zhang \& Zheng gratefully acknowledge the anonymous referee for reading our manuscript carefully and patiently.
Zhang \& Zheng gratefully acknowledge the anonymous referee for giving us constructive comments and 
suggestions to greatly improve our paper. Zhang gratefully acknowledges the kind grant support from NSFC-12173020. 
This paper has made use of the data from the SDSS projects, \url{http://www.sdss3.org/}, managed by the Astrophysical 
Research Consortium for the Participating Institutions of the SDSS-III Collaboration, and the data from MILES 
library \url{http://miles.iac.es/}. This paper has made use of the MPFIT package 
\url{https://pages.physics.wisc.edu/~craigm/idl/cmpfit.html}, and the emcee package 
\url{https://emcee.readthedocs.io/en/stable/}. This research has made use of the NASA/IPAC Extragalactic Database 
(NED, \url{http://ned.ipac.caltech.edu}). % which is operated by the California Institute 
%of Technology, under contract with the National Aeronautics and Space Administration.

\section*{Data Availability}
The data underlying this article will be shared on reasonable request to the corresponding author
(\href{mailto:aexueguang@qq.com}{aexueguang@qq.com}).

\label{lastpage}
\end{document}